\documentclass[reprint,aip,jcp ,superscriptaddress]{revtex4-2}

\usepackage{amssymb}
\usepackage{amsmath}
\usepackage{graphicx}
\usepackage{colortbl}
\usepackage{xcolor}
\usepackage[caption=false]{subfig}
\usepackage{varioref}
\usepackage[hidelinks]{hyperref}
\usepackage[nameinlink,capitalise]{cleveref}
\usepackage{siunitx}

\usepackage{comment}
\usepackage[english]{babel}

\newcommand{\dif}{\text{d}}
\newcommand{\Dif}{\mathcal{D}}
\newcommand{\mean}[1]{\left\langle #1 \right\rangle}
\newcommand{\MBR}{\text{MBR}}

\newcommand{\Obser}[1]{O \left[ #1\right]}

\newcommand{\sign}{\text{sign}}
\newcommand{\reg}{\vartheta}
\newcommand{\abs}[1]{\left\vert +1 \right\vert}

\date{March 2025}

\begin{document}

\title{Time reversal breaking of colloidal particles in cells}

\date{\today}
\author{Gabriel Knotz}
\affiliation{Institute for Theoretical Physics, University of Göttingen, 37077 Göttingen, Germany}
\author{Till M. Muenker}
\affiliation{Third Institute of Physics, University of Göttingen, 37077 Göttingen, Germany}
\author{Timo Betz}
\affiliation{Third Institute of Physics, University of Göttingen, 37077 Göttingen, Germany}
\author{Matthias Krüger}
\affiliation{Institute for Theoretical Physics, University of Göttingen, 37077 Göttingen, Germany}

\begin{abstract} 
    We investigate signatures of broken time reversal symmetry in stochastic trajectory data, employing the previously introduced three point correlation called mean back relaxation. We specifically investigate data from a simple driven model, as well as from colloidal particles within living or passivated biological cells.  
    Both in the model as well as in cell data, MBR detects broken time reversal symmetry, and furthermore, allows to determine relevant time and length scales of activity. For the cells, we show, by applying various drugs, that it is predominantly the presence of microtubules which is needed for a time reversal symmetry breaking. We employ a bound for entropy production, finding that it is in striking relation to previously determined active energies that quantify violation of the fluctuation dissipation theorem.         
\end{abstract}

\maketitle

\section{Introduction}
The study of non-equilibrium processes is a vibrant and growing field of research in physics as well as biophysics. A key challenge in this field is determination of signatures that distinguish equilibrium from non-equilibrium systems. Finding such signatures can be particularly challenging when dealing with highly coarse-grained dynamics, such as, e.g., colloidal particles in complex environments \cite{battle_broken_2016,santra_forces_2024}.
There are various established signatures. A standard method \cite{guo_2014, Turlier2016-dg, ahmed_active_2018,stoev_active_2025} is testing the validity of the fluctuation-dissipation theorem (FDT) \cite{kubo_fluctuation-dissipation_1966, agarwal_fluctuation-dissipation_1972,harada_equality_2005,speck_restoring_2006,baiesi_fluctuations_2009,kruger_fluctuation_2009,baiesi_update_2013}, whose violations are considered a signal for non-equilibrium. This procedure however requires measuring the system's mechanical response, for example, using optical or magnetic tweezers \cite{Vos2024-wo,Nishizawa2017-nb,Ebata2023-pi}. This can be experimentally challenging, as perturbations must be small to stay in the linear response regime. Another direct signature is the observation of probability currents as indicators of non-equilibrium processes \cite{zia_probability_2007, thapa_nonequilibrium_2024}, or the breakage of time reversal symmetry \cite{Mori22}. Formally, the latter is quantified by entropy production, for which various theorems exist \cite{evans_probability_1993,maes_time-reversal_2003,seifert_stochastic_2012}. Entropy production is typically hard to measure directly, and recent years have seen the developments of  bounds for it, e.g., in terms of moments of currents or time-antisymmetric observables \cite{barato_thermodynamic_2015,horowitz_thermodynamic_2020, hasegawa_fluctuation_2019, dieball_direct_2023, knotz_entropy_2024,seifert_stochastic_2019} or by other 
methods \cite{roldan_quantifying_2021,bacanu_inferring_2023,lynn_broken_2021}.

In this manuscript, we demonstrate that stochastic trajectories of Brownian particles immersed in living cells break time reversal symmetry, thereby demonstrating the non-equilibrium nature by purely passive observation \footnote{For simplicity, we take time reversal symmetry breaking as a measure of non-equilibrium, i.e., excluding cases that may break this symmetry by other means, e.g., magnetic fields \cite{kubo2012statistical}}. The investigation utilizes the mean back relaxation (MBR), a recently introduced observable \cite{muenker_accessing_2024,knotz_mean_2024,knotz_evaluating_2025} based on three time points. MBR contains a path antisymmetric part which is used to detect time reversal symmetry breaking, and which also allows to infer the length and time scales on which this symmetry is predominantly broken. Finally, we employ a previously derived bound for entropy production, thus providing a lower bound of entropy production in living cells. The so found bound is in qualitative agreement with previously determined active energies.
 We generalize the previously introduced random horse and cart model \cite{knotz_evaluating_2025} to a non-Gaussian version to allow for time reversal symmetry breaking. The model allows to illustrate the findings observed in cells.

The manuscript is organized as follows. In \cref{sec:MBR} we briefly introduce MBR. In \cref{sec:Model} we investigate a model system that exhibits time reversal breaking, which is detected by MBR, and we investigate how MBR can be used to determine the time and length scales of predominant breaking of time reversal symmetry. In \cref{sec:Cell} we apply MBR to cell data and demonstrate that it detects time reversal breaking for multiple different cell types. By investigating drugged cell types, we find that the time reversal breaking is primarily caused by processes associated with microtubules. Finally, in \cref{sec:bound}, we apply an entropy production bound and demonstrate qualitative agreement between the bound and a previously determined active energy.

\section{Mean Back Relaxation}
\label{sec:MBR}
We define mean back relaxation (MBR) for a stochastic variable $x(t)$ as the average of the negative ratio between the displacements $x(t) - x(0)$ and $x(0) - x(-\tau)$, with $\tau>0$, i.e., 
\begin{align}
  \notag\MBR(\tau,t,l)&= -\left\langle\frac{x(t) - x(0)}{x(0) - x(-\tau)} \vartheta_l(x(0) - x(-\tau))\right\rangle
  \\ 
  &= \notag\int \dif x_t \dif x_0 \dif x_{-\tau} \left( - \frac{x_t - x_0}{x_0 - x_{-\tau}} \right)  \\
   &             \times \vartheta_l(x_0 - x_{-\tau}) W_3(x_t,t;x_0,0;x_{-\tau},-\tau), 
    \label{eq:BR}
\end{align}
$W_3(x_{t_3},t_3;x_{t_2},t_2;x_{t_1},t_1)$ is the joint three-point probability of finding the particle at positions $x_{t_3},x_{t_2},x_{t_1}$ at times $t_3>t_2>t_1$. The factor $\vartheta_l(x(0) - x(-\tau)) \equiv \frac{\theta (|x(0) - x({-\tau})| - l)}{Z}$ with the Heaviside function $\theta(x)$, $Z= \mean{\theta \left(\left\vert x(0) - x(-\tau) \right\vert - l \right)}$ and a length $l>0$ avoids division by zero in \cref{eq:BR}. By construction, it is normalized to unity, i.e., $\mean{\vartheta_l(x(0) - x(-\tau))} = 1$, and the length $l$ is the smallest value of  $|x_{0} - x_{-\tau}|$ considered in \cref{eq:BR}, i.e., a cutoff length. 
We introduced a minus sign in \cref{eq:BR} to obtain a positive result for MBR in the case of back relaxation, i.e., when the displacements have opposite signs. MBR depends on the mentioned cut off length $l$, and the two times $t$ and $\tau$ at which the displacements in Eq.~\eqref{eq:BR} are evaluated.

MBR has been proven to be a marker of time reversal symmetry breaking \cite{knotz_mean_2024}: Under the  condition that the mean $\langle x\rangle$ exists and is approached in finite time, one has, for a process obeying detailed balance, \cite{knotz_mean_2024} 
\begin{align}
    \lim_{t \to \infty} \MBR(\tau,t,l) = \frac{1}{2},
    \label{eq:MBR-1-2}
\end{align}
independent of $\tau$ and $l$.

We have not been able to demonstrate  that the conditions for Eq.~\eqref{eq:MBR-1-2} are fulfilled for the particle in a cellular surrounding \cite{muenker_accessing_2024}, this is why we will not make use of Eq.~\eqref{eq:MBR-1-2} in this manuscript.  Instead, we will evaluate the time anti-symmetric part of MBR, which is distilled by taking the difference between MBR evaluated from forward and backward trajectories
\begin{align}
\begin{split}    \MBR_\text{anti}(\tau,t,l) = \left< - \frac{x(t+\tau) - x(\tau)}{x(\tau) - x(0)} \reg_l(x(\tau) - x(0)) \right.\\
    \left.+ \frac{x(0) - x(t)}{x(t) - x(t+\tau)} \reg_l(x(t) - x(t+\tau)) \right>.
    \end{split}\label{eq:MBR_anti}
\end{align}
Being the average of a path antisymmetric observable, $\MBR_\text{anti}$ must vanish for a time reversal symmetric process.  A finite value of $\MBR_\text{anti}$ thus signals breakage of time reversal symmetry. This will be explored for a toy model in \cref{sec:Model} and for experimental data obtained in cells in \cref{sec:Cell}. 
\section{Model: Discrete Random Horse and Cart}
\label{sec:Model}
\subsection{Model}

\begin{figure*}
    \centering
    \includegraphics[width=\linewidth]{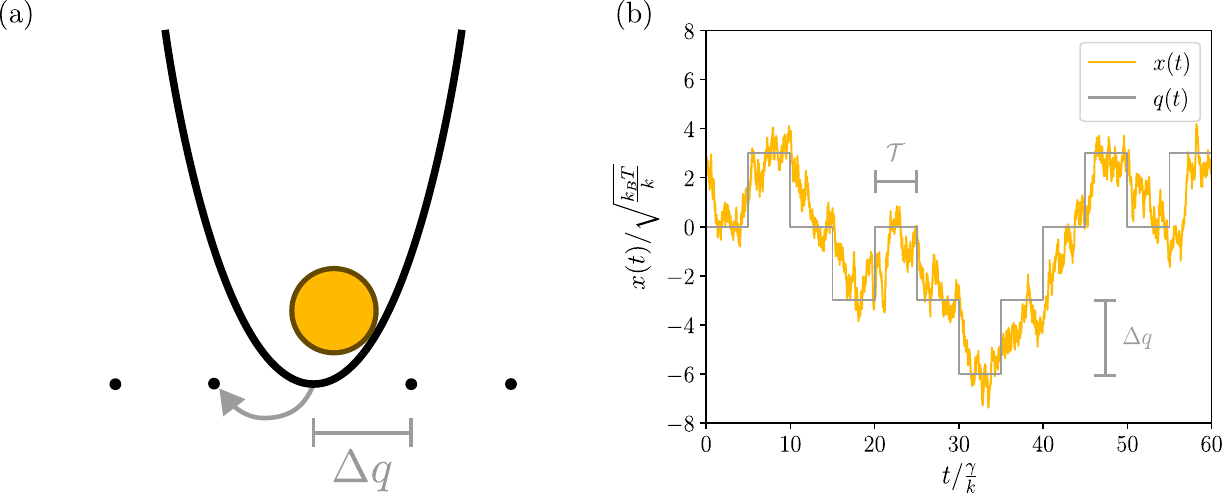}
    \caption{(a) Sketch of the dRHC model of \cref{eq:dRHC}. A Brownian particle is trapped in a harmonic potential whose center performs a discrete random walk with time intervals $\cal{T}$ and step length $\Delta q$. (b) Shows an example trajectory of $x$ and $q$ for $\Delta q = 3.0 \sqrt{\frac{k_B T}{k]}}$ and $\mathcal{T} = 5.0 \frac{\gamma}{k}$. This illustrates that $q$ performs discrete steps in space and time, while $x$ follows with a delay.}
    \label{fig:kp-model}
\end{figure*}

We start by investigating $\MBR_\text{anti}$ for a model system. In previous work \cite{muenker_accessing_2024, knotz_evaluating_2025}, we introduced the "random horse and cart" (RHC) model, set out to mimic the motor activity in living cells. The simplest version contains the particle $x$ of interest and a degree $q$ that mimics the role of an active environment, e.g., motor activity in biological cells. The equations of RHC read
\cite{muenker_accessing_2024},
\begin{subequations}\label{eq:model}
\begin{align}
    \dot x &= -\frac{k}{\gamma} (x - q) + \xi_x,\\
      \dot q &= \xi_q,
\end{align}
\end{subequations}
with white noises $\mean{\xi_i(t) \xi_j(t^\prime)} = 2D_i \delta(t-t^\prime) \delta_{ij}$ and $i,j\in\{x,q\}$. In Eq.~\eqref{eq:model}, the coupling between $x$ and $q$ is non-reciprocal, i.e., while the particle $x$ is subject to a force $-k(x-q)$ with a force constant $k$, the dynamics of $q$ is independent of $x$. $q$ may thus be seen as a (diffusing) motor that is so strong ("strong as a horse") that it does not feel the presence of the particle. The system of Eqs.~\eqref{eq:model} is a two dimensional Gaussian process. Due to the mentioned non-reciprocal coupling, it breaks detailed balance for any finite value of $D_q$ \cite{muenker_accessing_2024}. However, the stochastic process $x$ resulting from Eqs.~\eqref{eq:model} is a one dimensional Gaussian process, which does not break time reversal symmetry \cite{weiss_time-reversibility_1975, netz2023multipoint}. In other words, evaluating $\MBR_\text{anti}$ in Eq.~\eqref{eq:MBR_anti} for the process of Eq.~\eqref{eq:model} must yield a vanishing result.

\begin{figure*}
    \centering
    \includegraphics[width=\linewidth]{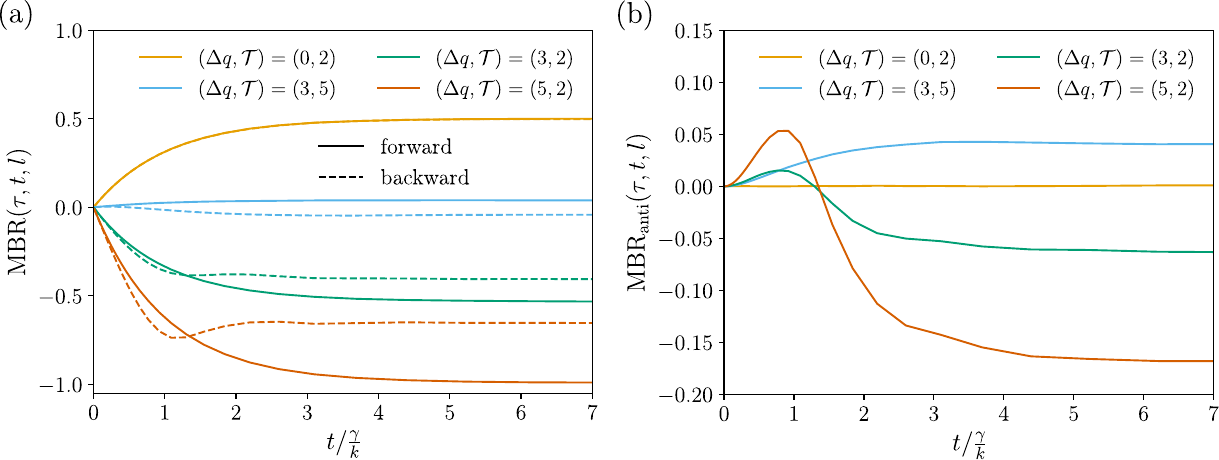}
    
    \caption{(a) MBR for the dRHC model of \cref{eq:dRHC} as a function of time $t$ for $\tau =  \frac{\gamma}{k}$,  $l= 0.01 \sqrt{\frac{k_B T}{k}}$ and 
$(\Delta q, {\cal T}) \in \left\lbrace (0,2), (3,5), (3,2), (5,2) \right\rbrace$
     with $\Delta q$ in units of $\sqrt{\frac{k_B T}{k}}$ and ${\cal T}$ in units of $\frac{\gamma}{k}$. Solid (dashed) lines: MBR from forward (backward) trajectories. Differences between solid and dashed signal time reversal symmetry breaking. Blue lines show the case of $\Delta q=0$ (equilibrium), and solid dashed are on top of each other. (b) $\MBR_\text{anti}$ of Eq.~\eqref{eq:MBR_anti} for the same parameters, i.e., difference between solid and dashed lines of panel a). 
    }
    \label{fig:kp-mbr}
\end{figure*}

In order to obtain a process that breaks time reversal symmetry, we thus render the model non-Gaussian, by making the  motor dynamics $q$ a discrete symmetric random walk with step length $\Delta q$ and time interval between steps of ${\cal T}$, see Fig.~\ref{fig:kp-model} for a sketch. The equations of the discrete random horse and cart model (dRHC) read for $t>0$
\begin{subequations}\label{eq:dRHC}
\begin{align}
     \dot x &= -\frac{k}{\gamma} ( x - q) + \xi_x\\
    q &= q_0 + \sum_{n=0}^{\text{floor}\left(\frac{t}{{\cal T}}\right)} \epsilon_n \Delta q
\end{align}
\end{subequations}
with $\mean{\xi_x(t) \xi_x(t^\prime)} = 2D_x \delta(t-t^\prime) $,  $\epsilon_n \in \left\lbrace-1,1 \right\rbrace$ with symmetric probabilities $p(\epsilon_n = 1) = p(\epsilon_n = -1)=\frac{1}{2}$. The dRHC model is sketched in \cref{fig:kp-model} where we also show an example trajectory. 

As mentioned, dRHC differs in one  crucial aspect from RHC in Eq.~\eqref{eq:model}: The resulting process $x$ is non-Gaussian, and potentially allows detection of time reversal symmetry breaking. Additionally, the active driving mechanism in dRHC introduces a driving length scale, allowing to investigate whether MBR allows detection of this scale. 
\subsection{MBR  and time reversal breaking}
\Cref{fig:kp-mbr} a) shows MBR  from both forward and reversed trajectories from dRHC, obtained from numerical solution (simulation) of Eq.~\eqref{eq:dRHC}. The blue curve shows the case of zero step length $\Delta q$, i.e., the case of a Brownian particle in  a stationary potential. In this case, the process of $x$ obeys time reversal symmetry, and MBR approaches $\frac{1}{2}$ \cite{muenker_accessing_2024}. Furthermore, as this case is time reversal symmetric, MBR evaluated from forward and MBR from backward trajectories are identical -- the dashed and solid lines are on top of each other. However, for finite step lengths $\Delta q$ (purple and red curves, see labels), time reversal symmetry is broken: MBR evaluated from forward and backward  trajectories differ. This difference grows with the step length $\Delta q$.

\Cref{fig:kp-mbr} b) shows $\MBR_\text{anti}$ of Eq.~\eqref{eq:MBR_anti}, i.e.,  the difference between solid and dashed curves in Fig.~\ref{fig:kp-mbr}, for the same parameters with the same color code. This graph highlights the breaking of time reversal symmetry for the cases with finite $\Delta q$. 

\subsection{Dependence on length and time scales}

The model of Eq.~\eqref{eq:dRHC} is driven by the active process $q$ with inherent time scale $\cal T$ and length scale $\Delta q$. Can we extract these scales in MBR evaluated for $x$? We will discuss this question in this subsection, to be later able to address it in case of living cells.

\begin{figure}
    \centering
    \includegraphics[width=\linewidth]{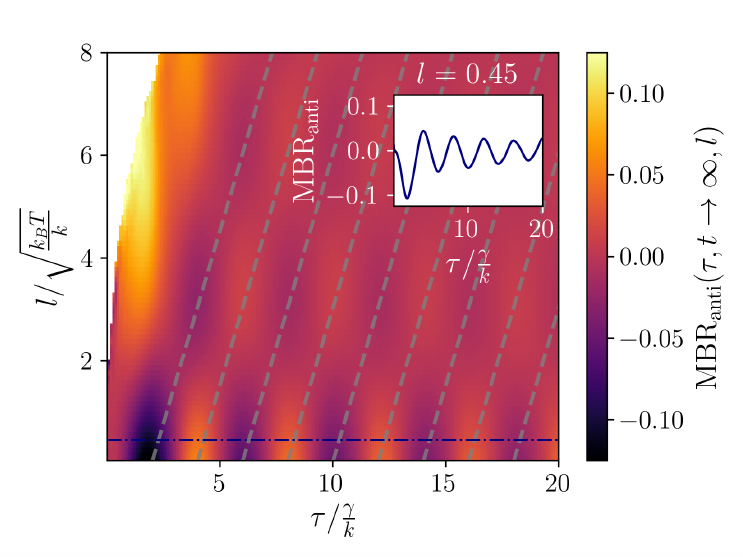}
    \caption{Anti-symmetric part of MBR, i.e., $\MBR_\text{anti}$ of Eq.~\eqref{eq:MBR_anti}, for trajectories from dRHC model of \cref{eq:dRHC} as functions of $\tau$ and $l$. Model parameters are  $\Delta q = 3 \sqrt{\frac{k_B T}{k}}$ and ${\cal T} = 2.0 \frac{\gamma}{k}$. Gray lines are guides to the eye, and they follow  \cref{eq:kp-slope_grid}. Horizontal distance between neighboring dashed lines is ${\cal T}$, vertical distance is $\Delta q$. This allows extraction of the active time and length scale. Inset shows $\MBR_\text{anti}$ as a function of $\tau$ for $l=0.45 \sqrt{\frac{k_B T}{k}}$ (indicated by dash-dotted line in main graph), illustrating the oscillatory nature.}
    \label{fig:kp-LT}
\end{figure}

Figure \ref{fig:kp-LT} presents the anti-symmetric component of long-time MBR, i.e., $\MBR_\text{anti}(\tau,t\to\infty,l)$ of Eq.~\eqref{eq:MBR_anti}, for the dRHC   model of \cref{eq:dRHC}, for fixed model parameters, as a color map across the MBR-parameters $\tau$ and $l$. This visualization reveals a striking observation: The map shows a periodic structure of alternating maxima and minima. This periodicity is visible in both the $\tau$ and $l$ directions. Even more striking: The distance between successive peaks along the $\tau$ direction corresponds very precisely to the interval of the non-equilibrium driving, namely ${\cal T}$: The dashed gray lines, added as guide to the eye, have exactly horizontal distances of ${\cal T}$.

The vertical separation observed in \cref{fig:kp-LT} corresponds, again very precisely, to $\Delta q$: The vertical distance between neighboring grey dashed lines is given by $\Delta q$. Precisely, the gray dotted lines follow 
\begin{align}
        l(\tau) = \frac{\Delta q}{{\cal T}} \left( \tau - n \cdot {\cal T} \right)
        \label{eq:kp-slope_grid}
    \end{align}
with $n \in \mathbb{N}$.

The striking finding of \cref{fig:kp-LT}, with Eq.~\eqref{eq:kp-slope_grid}, is a strong indication that the length and time scales of non-equilibrium mechanisms can indeed be distilled via MBR.

\begin{figure}
    \centering
    \includegraphics[width=\linewidth]{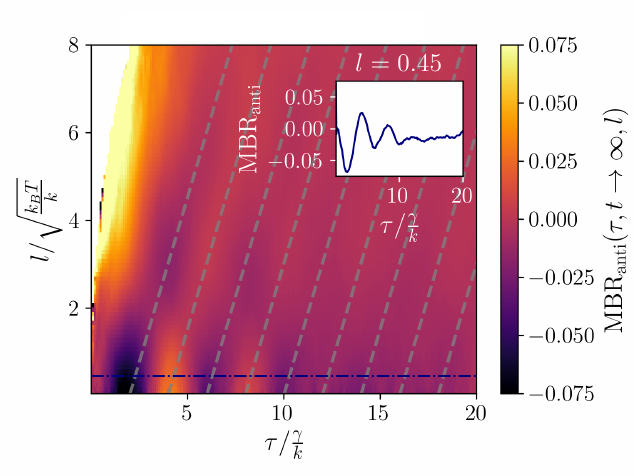}
    \caption{Anti-symmetric part of MBR, i.e., $\MBR_\text{anti}$ of Eq.~\eqref{eq:MBR_anti}, for dRHC of \cref{eq:dRHC}, with all parameters identical to \cref{fig:kp-LT}, but with Gaussian distributed step intervals with  standard deviation $\sigma_t/{\cal T} = 1/4$.
    Gray lines are guides to the eye, and they follow  \cref{eq:kp-slope_grid}, i.e., horizontal distance between neighboring dashed lines is ${\cal T}$, vertical distance is $\Delta q$. Inset shows $\MBR_\text{anti}$ as a function of $\tau$ for $l=0.45 \sqrt{\frac{k_B T}{k}}$ (indicated by dash-dotted line in main graph), illustrating the decaying oscillatory nature.  }
    \label{fig:kpg-LT}
\end{figure}

The discrete random horse and cart model has a sharp step interval of ${\cal T}$, which gives rise to the sharp and pronounced features of \cref{fig:kp-LT}. To test the robustness of this finding against some randomness in the step interval, we now take the jump interval from a Gaussian distribution with mean ${\cal T}$ and a variance of $\sigma_t$. 
\cref{fig:kpg-LT}  mirrors \cref{fig:kp-LT} in all parameters, except for the randomly drawn step interval with a value of ${\cal T}/\sigma_t = 4$. The overall structure of the figure is similar to \cref{fig:kp-LT}, i.e., the oscillatory structure quantified by \cref{eq:kp-slope_grid} is visible. However, as may be expected, the structure is more washed out, and the amplitude of the peaks diminishes with larger values of $\tau$ and $l$. We conclude that, as expected,  variability in the driving process makes it harder to extract length and time scales via MBR. 

\section{Trajectories from living cells}

\begin{figure*}
 \centering
 \includegraphics[width=\linewidth]{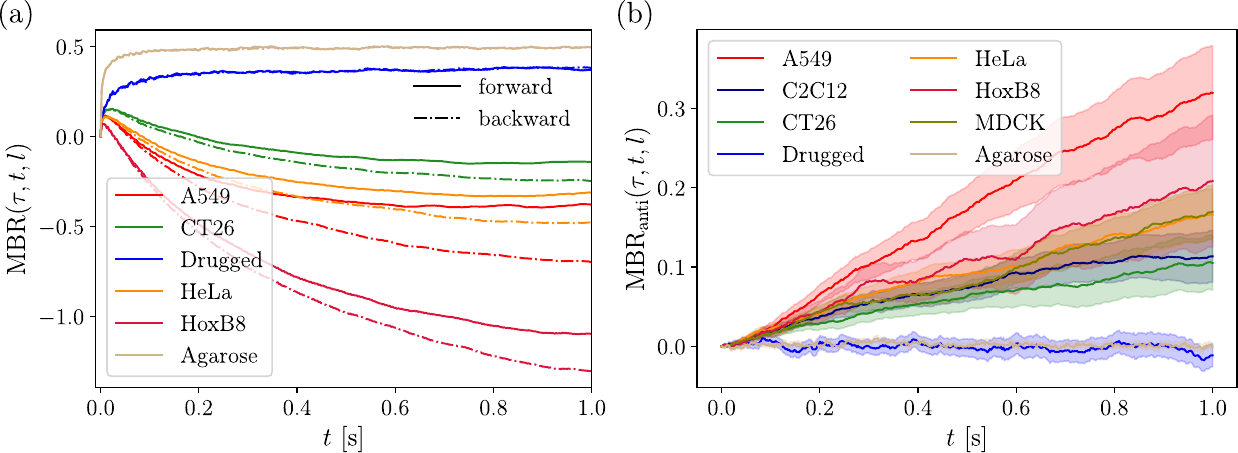}
     \caption{(a) MBR for various cell types, as labelled, for $\tau = \SI{0.047}{\second}$ and $l = \SI{0.002}{\micro\meter}$. Solid lines show MBR evaluated for time-forward trajectories and dotted lines for time-backward trajectories. For the wild-type cells, the solid and dotted lines disagree, indicating a breakage of time reversal symmetry. For the drug-treated HeLa cells, there is no significant difference between forward and backward curves. (b) Anti-symmetric part of MBR of Eq.~\eqref{eq:MBR_anti}, i.e., difference between respective dashed and solid lines in (a). Colored envelopes give the standard deviation of a bootstrap distribution (4000 samples). a) and b) also show MBR from trajectories of a probe particle in the purely passive medium of aragose, as a reference.} 
     \label{fig:MBR_cells}
 \end{figure*}

In this section, we analyze trajectories of probe particles in living cells \cite{muenker_accessing_2024,knotz_entropy_2024}, using the tools developed and tested for the dRHC model in \cref{sec:Model}. The cell types used are human epithelial lung cancer cells (A549), mouse metastatic lung cancer cells (CT26), mouse immune cells derived from immortalized bone marrow cells where differentiation is controlled by repressing the HoxB8 gene (HoxB8) and human cervical cancer cells untreated (HeLa), as well as HeLa cells treated with both, cytochalasin B and nocodazole, to depolymerize the actin and the microtubule network respectively (drugged). 
\label{sec:Cell}
\subsection{Time reversal breaking}
We investigated MBR curves from trajectories obtained in living wild type and passivated cells in previous works \cite{muenker_accessing_2024,knotz_evaluating_2025}. We observed astonishing relations between MBR and effective energies, the latter quantifying breakage of the fluctuation dissipation theorem \cite{muenker_accessing_2024}. However, the breakage of detailed balance of trajectories recorded in cells has, to our knowledge not been demonstrated.    

Figure~\ref{fig:MBR_cells} a) shows MBR evaluated for various cell types, as a function of time. We observe clear and pronounced differences between MBR evaluated from forward and backward trajectories, demonstrating a breaking of time reversal symmetry. In contrast, for the drug-treated HeLa cells, there is no difference between forward and back cases, i.e., breakage of time reversal symmetry cannot be detected via MBR in these cells. 

 Figure~\ref{fig:MBR_cells} b) shows the difference between the curves in a), namely the antisymmetric part defined in \cref{eq:MBR_anti}. In this curve, we also provide error areas, corresponding to the standard deviation obtained by bootstrap analysis. The finiteness of $\MBR_\text{anti}$ is thus statistically significant, i.e., larger than error bars, and we can claim to have demonstrated or detected the breakage of detailed balance in these living cells. For the drugged cell, the value of $\MBR_\text{anti}$ includes zero within error, and no significant breakage of time reversal symmetry can be claimed.
 
 The mentioned drugs disrupt actin and microtubules, reducing cytoskeletal motor activity, which likely explains the absent time reversal breaking in these cells. 

 \cref{fig:MBR_cells} a) and b) also include trajectories from probe particles in agarose, a purely passive medium. the behavior of MBR is similar to the one in the drugged cell, i.e., no breakage of time reversal symmetry is found.

\subsection{Dependence on length and time scales}

\begin{figure*}
    \centering
    \includegraphics[width=\linewidth]{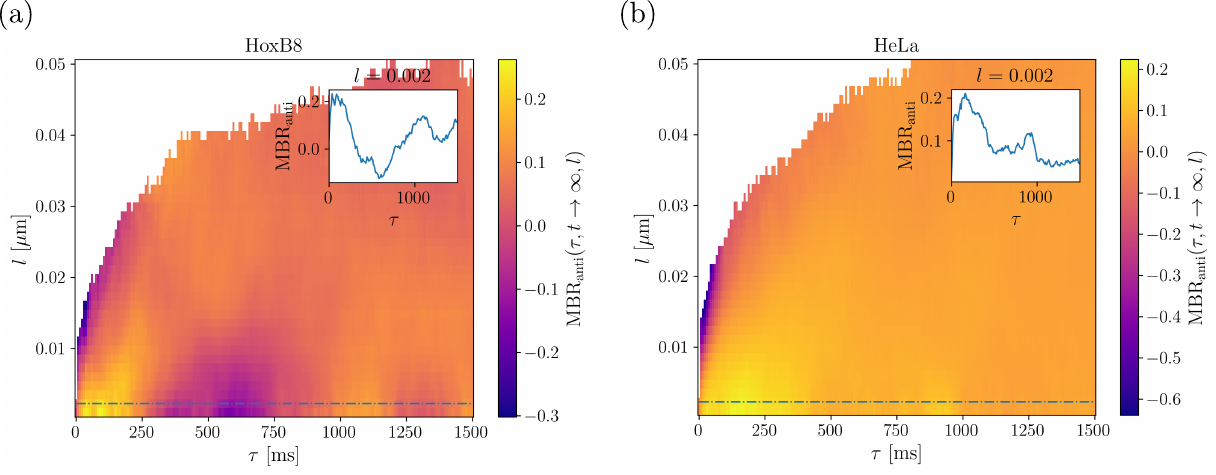}
        \caption{$\MBR_\text{anti}$ at $t=\SI{1}{\second}$ as functions of $\tau$ and $l$ for (a) HoxB8 and (b) HeLa cells. The periodic structure of \cref{fig:kp-LT} is notable, however  washed out. From the periodicity, we estimate the time and length  scales to roughly $ \SI{500}{\milli \second}$ and $ \SI{20}{\nano \meter}$, respectively, in both cell types. Insets show $\MBR_\text{anti}$ as functions of $\tau$ for fixed $l$ (i.e., along the dash-dotted lines in the main graphs), indicating another timescale at around $ \SI{100}{\milli \second}$ or below. }
        \label{fig:Hox-mbr-LT}
\end{figure*}
In \cref{fig:Hox-mbr-LT} we show $\MBR_\text{anti}$ at $t=\SI{1}{\second}$, as functions of $\tau$ and $l$, for HoxB8 (a) and HeLa (b) cells, in close resemblance of \cref{fig:kp-LT}. Both figures show peak structures, indeed similar to \cref{fig:kp-LT}, however washed out, the more so for the case of HeLa. Comparing to \cref{fig:kp-LT}, this periodicity  suggests active length and time scales of roughly $ \SI{500}{\milli \second}$ and $ \SI{20}{\nano \meter}$, respectively, in both cell types. 
The insets in \cref{fig:Hox-mbr-LT} show $\MBR_\text{anti}$ as a function of $\tau$ for a fixed $l$, indicating another peak at around $ \SI{100}{\milli \second}$ or below. \Cref{fig:cells_LT} in \cref{app:plots} shows the same for the other untreated cells where we also oberve washed out peaks.
The particle trajectories analyzed here report the motion of phagocytosed polystyrene beads with a diameter of 
$1 \mu m$. At the time of recording, these non-degradable particles represent membrane-covered rigid objects that are stored in the cell similarly to lysosomes. Lysosomes are typically transported inside cells by microtubule-based motors, such as kinesins or dynein, which can operate in a tug-of-war configuration.

Given the complexity of the cytosol and our observation that these particles do not move processively once they are close to their destination, characteristic timescales on the order of 100–500 ms are very reasonable \cite{lipowsky_life_2005,deguchi_direct_2023}. Furthermore, kinesin motors are known to exhibit step sizes of approximately 8 nm, while dynein can often skip binding sites on the microtubule lattice, leading to step lengths of up to ~32 nm.

Comparison with the observed cellular data therefore suggests that kinesins, dyneins, or a combination of both could represent the underlying molecular origin of the observed MBR structure. Overall, the approach predicts that dynein is the dominant source of non-equilibrium activity, as it better matches the observed length and timescales. This suggests that polymerization of microtubules should have a stronger effect on the observed peaks than depolymerizing actin \cite{muenker_intracellular_2024}.

The maps for both cells however show a feature not visible for the dRHC model of  \cref{fig:kp-LT}: On the left hand side of the maps, there is an area of pronounced negative $\MBR_\text{anti}$ values. This will be discussed in the next subsection, where we show these on time logarithmic axes. 

\subsection{Drugging HeLa Cells} 
\begin{figure*}
    \centering
    \includegraphics[width=\linewidth]{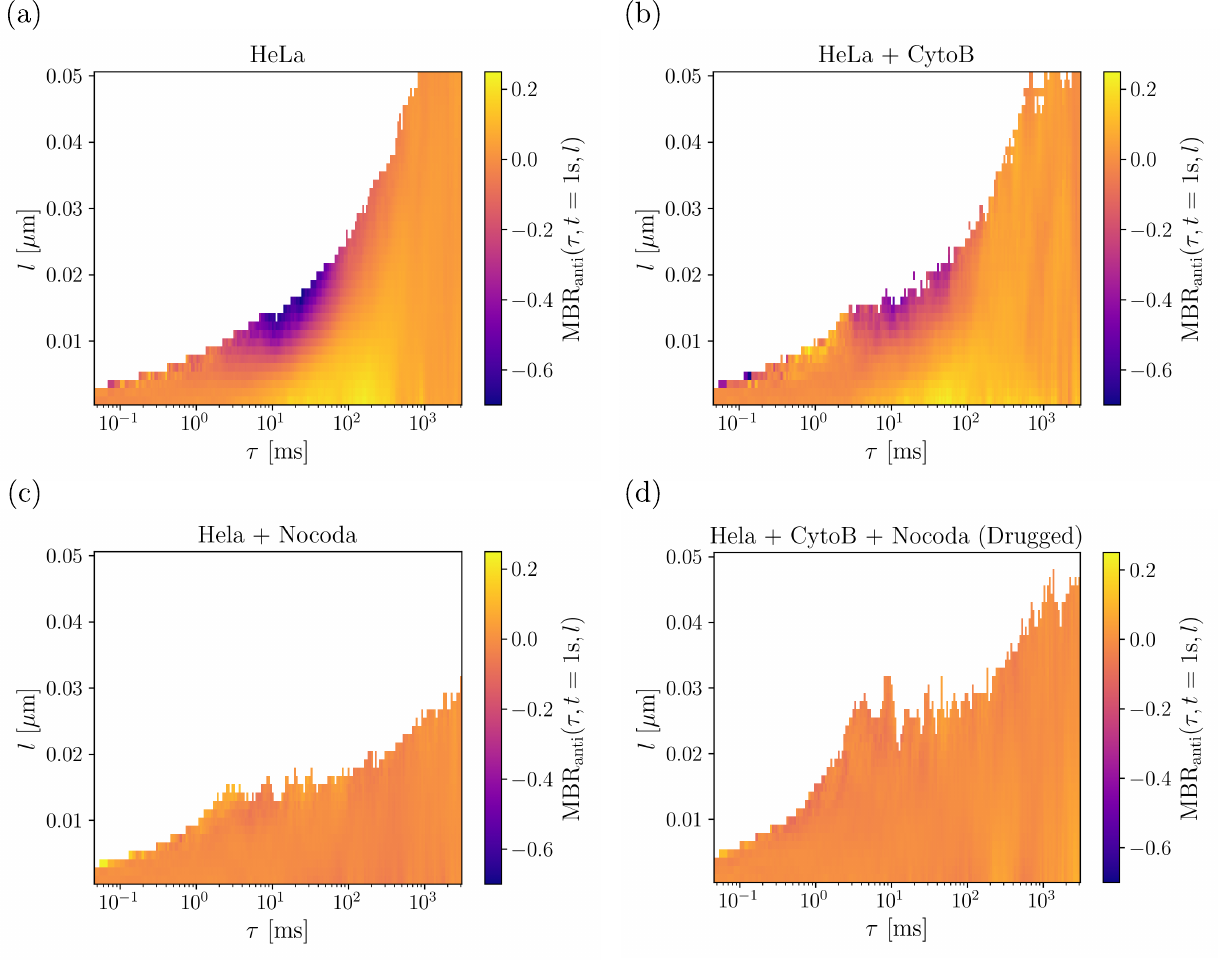}
    \caption{Anti-symmetric part, $\MBR_\text{anti}$ at  $t=\SI{1}{\second}$ as functions of $\tau$ and $l$ for HeLa cells treated with different drugs. In white areas insufficient amount of events was recorded to generate a reliable average value, which is at the base of the MBR approach. (a) Wild type, i.e., undrugged cell. (b) Drugged with Cytochalasin B (CytoB), a drug that removes actin. (c) Drugged with Nocodazole (Nocoda), a drug that disrupts microtubules.  (d) Drugged with both CytoB and Nocoda. Case d) is referred to as the "drugged case" of  \cref{fig:MBR_cells}. }
    \label{fig:HeLa-CytoB-Nocoda}
\end{figure*}
In \cref{fig:HeLa-CytoB-Nocoda}, we show the anti-symmetric part of MBR at $t=\SI{1}{\second}$ for multiple HeLa cell variants, each as functions of $\tau$ and $l$. \cref{fig:HeLa-CytoB-Nocoda} is similar as \cref{fig:Hox-mbr-LT}, but using an logarithmic $\tau$ axis, and comparing different stages of drugging. \cref{fig:HeLa-CytoB-Nocoda}a) shows the wild type case, i.e., this map shows identical data as \cref{fig:Hox-mbr-LT}b), except presented on a logarithmic $\tau$ axis. As expected, the periodicity discussed around Figs.~\ref{fig:kp-LT} and \ref{fig:Hox-mbr-LT} is not evident in the logarithmic representation. However, this representation emphasized the strongest antisymmetric signal that is found just at the border of the detectable regime. This signal corresponds to a length scale of $10-20 \si{\nano\meter}$ and a timescale of $5-50\si{\milli\second}$. Indeed these scale correspond perfectly to the expected signal from dynein motor proteins.  

As dynein is a microtubule based motor protein, we systematically disrupted either microtubulues, actin filaments, or both. If our hypothesis that the anti-symmetric signature indeed corresponds to dynein, then we expect a strong effect when depolymerizing microtubules, and a less strong effect when disrupting the actin cytoskeleton. Indeed as shown in 
\cref{fig:HeLa-CytoB-Nocoda}b), HeLa cell with actin removed by drugging with CytoB still have a persisting time reversal breaking. Although the amplitude or extent of reversal breaking is slightly reduced, this implies that actin is not crucial for the processes breaking time reversal. In contrast, when depolymerizing the microtubules using Nocodazole (\cref{fig:HeLa-CytoB-Nocoda} c) no time reversal breaking is notable in the MBR, implying that microtubules are important for the cell's active processes. This is further confirmed when disrupting both, actin filaments and microtubules (\cref{fig:HeLa-CytoB-Nocoda} d) which resembles the situation of only removing microtubulues. This analysis suggests that the time reversal breaking is mainly linked to active processes related to the microtubules, and is consistent with the hypothesis that the main force generating motor protein is dynein.

\section{Bound for entropy production}\label{sec:bound}
\subsection{General: Entropy bound from MBR} 

Can we quantify the breaking of time reversal symmetry in the investigated cells? To answer this question we consider a path observable $\Obser{\omega}$ that depends on the path $\omega$ in phase space, with path probability $p[\omega]$. The average of this observable is formally given by the sum over paths \cite{altland_condensed_2010}
  $  \mean{O} = \int \Dif \omega ~ \Obser{\omega} p[\omega]$. 
To quantify the mentioned path reversal properties, we resort to the stochastic change in entropy defined as the log ratio of path probabilities \cite{Maes2003, seifert_entropy_2005, fischer_free_2020}
\begin{align}
    s_{} &= \log \frac{p[\omega]}{p[\theta \omega]}.
    \label{eq:def-ent}
\end{align}
We use notation for path reversal, $\theta \omega$, including reversal of time as well as of kinematic reversal of momenta \cite{spinney_entropy_2012}. For simplicity, we will in the following refer to $s$ as the entropy production (formally per Boltzmann constant $k_B$), despite some caveats regarding this term \footnote{The entropy defined in Eq.~\eqref{eq:def-ent} corresponds to the total change in entropy in overdamped stationary systems. In underdamped or non-stationary systems the boundary terms differ \cite{fischer_free_2020}.}. 

We will use $s$ in Eq.~\eqref{eq:def-ent} as a quantifier of broken time reversal symmetry and activity in the investigated cells.
Specifically, we will make use of the following inequality, or bound \cite{knotz_entropy_2024}
\begin{align}
    \mean{s} \geq \mean{O_a} \log \left( \frac{1 + \mean{O_a}}{1 - \mean{O_a}} \right).
    \label{eq:s-bound}
\end{align}
Eq.~\eqref{eq:s-bound} provides a bound for entropy production $\mean{s}$ based on any anti-symmetric observable $O_a[\omega] = -O_a[\theta \omega]$ with $\vert O_a \vert \leq 1$. Eq.~\eqref{eq:s-bound} can be used with $\MBR_\text{anti}$ as a basic anti-symmetric observable, as shown below, where we focus on the sign of $\MBR_\text{anti}$. Furthermore, Eq.~\eqref{eq:s-bound} is based on a fluctuation theorem \cite{knotz_entropy_2024}, and hence enjoys broad validity.    

Specifically, we express  $O_a$ in terms of the antisymmetric part of MBR,
\begin{align}
\begin{split}   O_a(\tau,t,l)  = \sign\left( - \frac{x(t+\tau) - x(\tau)}{x(\tau) - x(0)} \reg_l(x(\tau) - x(0)) \right.\\
    \left.+ \frac{x(0) - x(t)}{x(t) - x(t+\tau)} \reg_l(x(t) - x(t+\tau)) \right).
    \end{split}\label{eq:o-sig}
\end{align}
The antisymmetric observable chosen in Eq.~\eqref{eq:o-sig} is based on the observable averaged over in $\MBR_\text{anti}$ in Eq.~\eqref{eq:MBR_anti}, with an additional operation of $\sign$, to ensure $\vert O_a \vert \leq 1$. The parameters $\tau$, $l$ and $t$ may now be varied in Eq.~\eqref{eq:o-sig} to maximize the bound in \cref{eq:s-bound}.

\subsection{Entropy bound from dRHC model}\label{sec:EBM}
\begin{figure}
    \centering
    \includegraphics[width=\linewidth]{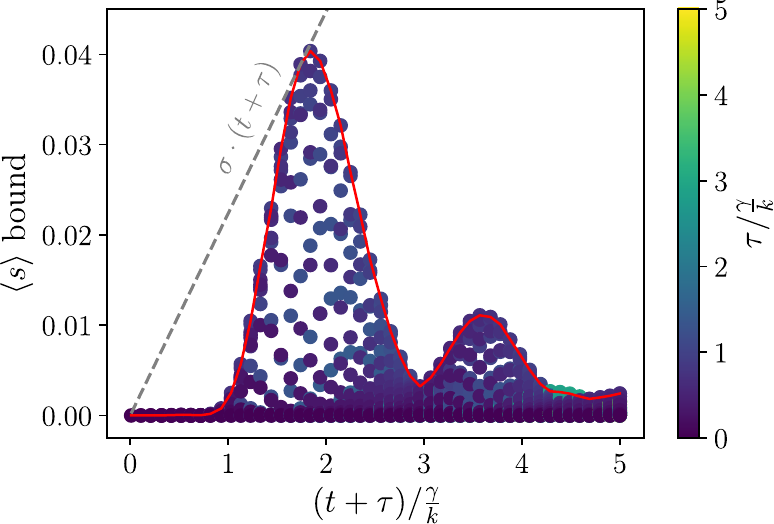}
    \caption{Entropy bound, \cref{eq:s-bound}, using \cref{eq:o-sig} for the dRHC model of \cref{eq:dRHC} with $l= 0.05 \sqrt{\frac{k_B T}{k}}, \Delta q = 5.0 \sqrt{\frac{k_B T}{k}}$ and $\mathcal{T} = 1.0 \frac{\gamma}{k}$. Each point marks a distinct  combination of $t$ and $\tau$ adding up to $t+\tau$ on the $x$-axis, with $\tau$ as color coded. Red line shows the maximum of all points as a guide to the eye. Gray dotted line shows the largest bound for entropy production rate $\sigma$, with a slope of $\approx 0.022 \frac{k}{\gamma}$. }
    \label{fig:ent_bound_model}
\end{figure}
\Cref{fig:ent_bound_model} shows the right hand side of \cref{eq:s-bound} using the observable \cref{eq:o-sig} applied to the dRHC model for fixed model parameters $\Delta q$ and $\cal T$ as given. The graph shows a color map as functions of $\tau$ and $t+\tau$, for a fixed value of $l = 0.05 \sqrt{\frac{k_B T}{k}}$, motivated by \cref{fig:kp-LT}.

\Cref{fig:ent_bound_model}  reveals that the total trajectory length ($t+\tau$) largely dictates the bound on $\mean{s(t+\tau)}$, with multiple $(t, \tau)$ combinations yielding the same bound. 
Plotting \cref{fig:ent_bound_model} as a function of $t+\tau$ lies in extraction of entropy production {\it rate} $\sigma$. As the process is stationary, we may assume a constant rate $\sigma$ that yields its bound as 

\begin{align}
    \sigma \geq \max_{t,\tau} \frac{1}{t+\tau} \mean{O_a(\tau,t,l)} \log \left( \frac{1 + \mean{O_a(\tau,t,l)}}{1 - \mean{O_a(\tau,t,l)}} \right).
\end{align}
The largest bound for $\sigma$ is thus obtained from drawing a line through the origin with the smallest slope that bounds all points in \cref{fig:ent_bound_model} from above. This line is shown as a dashed line in \cref{fig:ent_bound_model}. With this procedure,  we obtain for the model parameters used in  \cref{fig:ent_bound_model} a bound of $\sigma_\text{bound} \approx 0.022 \frac{k}{\gamma}$.

To compare the bound with the actual entropy production rate, we numerically compute the work rate performed by the degree $q$ on $x$, i.e. \cite{seifert_stochastic_2012}, 
\begin{align}
    \sigma_{\rm work}  &= \frac{1}{k_B T} \lim_{t \to \infty} \frac{1}{t} \int_0^t F(x,q) \circ \dif x(s) \notag\\
    &= \frac{1}{k_B T} \lim_{t \to \infty} \frac{1}{t} \int_0^t -k(x(s) - q(s)) \circ \dif x(s)    \label{eq:model_entropy_production}
\end{align}
where $\circ\, \dif x(a)$ indicates that the integral is taken with Stratonovic convention. This yields, for the same model parameters $\sigma _{\rm work}\approx 12.265 \frac{k}{\gamma}$, significantly larger than the estimated bound for $\sigma$. While interesting, this comparison may, at first sight, be considered not very useful, as $\sigma_{\rm work}$ in \cref{eq:model_entropy_production} yields the work done, while $\sigma$ estimated from \cref{fig:ent_bound_model} is the entropy production from the trajectory of $x$. Naturally, the latter is smaller, which may also be understood from the above made statements: If the process  $x$ was Gaussian -- as, e.g., in the RHC model of Eq.~\eqref{eq:model} -- the estimate of $\sigma$ from $x$ would yield zero, while the work done in  \cref{eq:model_entropy_production} would still be finite. It is the non-Gaussianity in $x$ which is needed for a finite estimate from $x$. 

This being said, the comparison is useful, because for the cells investigated below, we face the same issue: We estimate entropy production from a single degree $x$, which naturally highly underestimates the total number of energetic processes in the cell.     

Finally, we aim to point out that the bound obtained from \cref{fig:ent_bound_model} naturally underestimates the entropy production of the process $x$ -- by how much is hard to say, because finding the true value is difficult. 
 
\subsection{Cells}

\begin{figure}
    \centering
    \includegraphics[width=\linewidth]{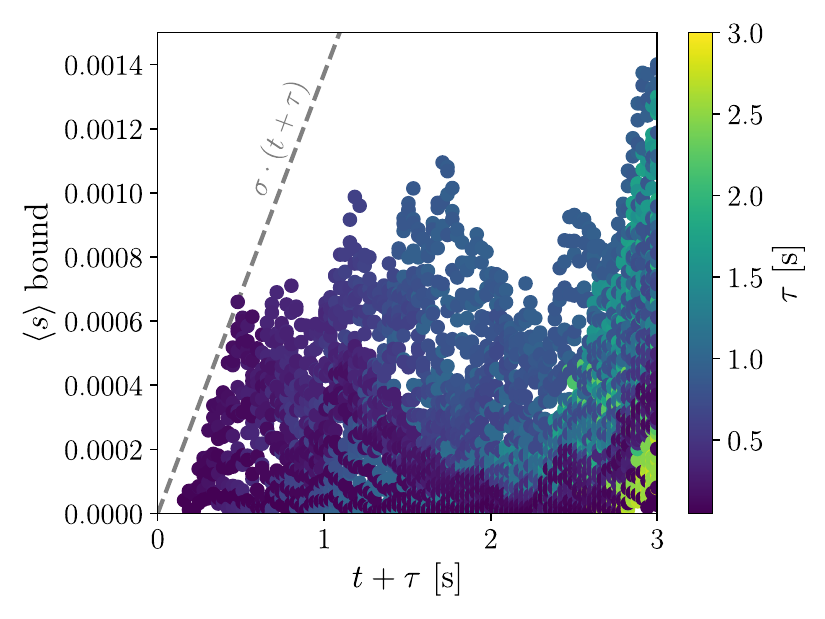}
    \caption{Entropy bound, \cref{eq:s-bound}, using \cref{eq:o-sig} for HeLa cells with $l = \SI{0.002}{\micro\meter}$. Each point marks a distinct  combination of $t$ and $\tau$ adding up to $t+\tau$ on the $x$-axis, with $\tau$ as color coded. Gray dotted line shows the best bound for entropy production rate $\sigma$, with a slope of $\approx 0.0014 \frac{1}{\SI{}{\second}}$. }
    \label{fig:s-bound}
\end{figure}
We now turn to the same analysis in cell data. \Cref{fig:s-bound} shows the right hand side of  \cref{eq:s-bound} using the observable in Eq.~\eqref{eq:o-sig}, as functions of combinations of $t$ and $\tau$ for HeLa cells. We choose a value of $l = \SI{0.002}{\micro\meter}$, because, according to \cref{fig:Hox-mbr-LT,fig:HeLa-CytoB-Nocoda}, this value -- which is on the edge of experimental resolution -- yields large time reversal symmetry breaking. As for the model system we see multiple combinations of $(t,\tau)$ yielding the same bound $\mean{s(t+\tau)}$. Following the procedure as in \cref{fig:ent_bound_model} yields a bound for entropy production rate $\sigma$. 

\begin{figure}
    \centering
    \includegraphics[width=\linewidth]{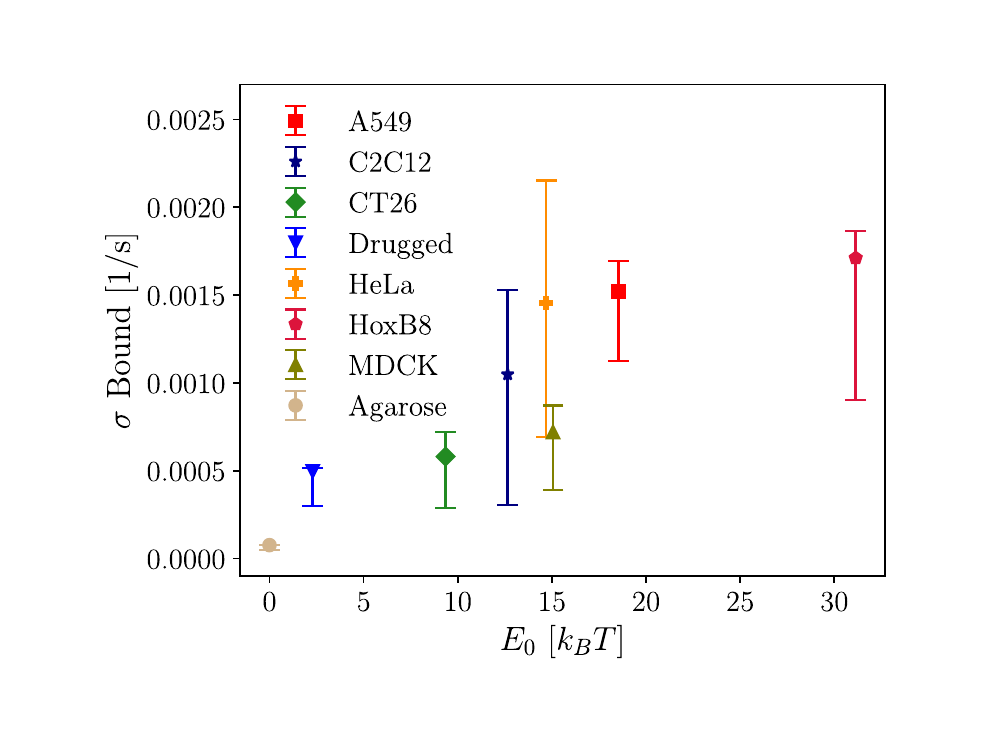}
    \caption{Bound for entropy production rate $\sigma$ for different cell types and agarose, as extracted from procedures in \cref{fig:s-bound}. Error bars are obtained via bias corrected and accelerated bootstrap (4000 samples) with a confidence interval of 95\%. $x$ axis shows the effective energy $E_0$, which was determined in Ref.~\onlinecite{muenker_accessing_2024}, and which quantifies violation of  FDT. }
    \label{fig:sigma_bound}
\end{figure}

\Cref{fig:sigma_bound} shows the so obtained bound for $\sigma$ for all cells investigated. As control measurement, we also show the bound calculated from trajectories of colloidal particle in agarose, a purely passive fluid. Indeed these particles yield the lowest bound for $\sigma$, thus providing  an idea of the accuracy of this method (see discussion in \cref{sec:discussion} below). For all values, errorbars show the 95\% confidence interval of a bias-corrected and accelerated bootstrap. We can see that the error for all cell types is rather large, which may indicate significant variability across different cells of the same cell type, consistent with measurements of active energy in these cells \cite{muenker_intracellular_2024}.

The values obtained for entropy production rate are small, and they are not to be understood as the entropy production rate of the cell, see also the discussion for the dRHC  model in \cref{sec:EBM}. As mentioned there, the reason for this smallness lies in the fact that only a single, one-dimensional degree is observed, and many other important degrees of cell dynamics are not. We remind that if such a single degree is Gaussian, it shows no entropy production at all, i.e., it can only be non-Gaussian features that contribute to $\sigma$. This shows the difficulty in estimating entropy production from a suspended particle. Further, Eq.~\eqref{eq:s-bound} is a bound, and Eq.~\eqref{eq:o-sig} is not the optimal observable (which is the signum of entropy production \cite{knotz_entropy_2024}). This implies that the entropy production of the single degree  is also likely underestimated in \cref{fig:sigma_bound}.   

Despite these difficulties, do the values of \cref{fig:sigma_bound} have a relative meaning between cells? 
To compare with another activity measure, we recall the effective energy, which quantifies the violation of FDT and was determined for the investigated cells in Refs.~\cite{muenker_intracellular_2024,muenker_accessing_2024}, using an optical tweezer setup. It is defined as the ratio \cite{martin_comparison_2001,ahmed_active_2018,narinder_time_2025}
\begin{align}
    E_\text{Eff}(\omega) = \frac{\omega C(\omega)}{2\chi^{\prime\prime}(\omega)},
    \label{eq:Eeff}
\end{align}
with the power spectral density $C(\omega)$ of the degree $x$ and the imaginary part of the response function $\chi^{\prime\prime}(\omega)$ of this degree. For the cells investigated here, it was observed that the effective energy follows a power law, i.e., $ E_\text{Eff}(\omega)\approx k_B T + E_0 \frac{\omega_0}{\omega}$. The prefactor $E_0$, determined at $\omega_0 = \SI{1}{\hertz}$, depends on the cell under investigation \cite{muenker_intracellular_2024}, and we take it as a measure of the cell's activity. 

\Cref{fig:sigma_bound} shows these values of $E_0$ on the $x$-axis, i.e., the cells are ordered from left to right according to the measured values of $E_0$. This yields monotonically growing values of $\sigma$, except for MDCK cells, whose value of $\sigma$ appears too small relative to the found effective energy. Given the uncertainty in both measures, the curve demonstrates convincingly that the two independent ways of quantifying the cell's activity yield a similar trend and hierarchy.

\section{Discussion}
\label{sec:discussion}
We investigated the breaking of time reversal symmetry in one dimensional stochastic trajectories, both from a driven model as well as from probe particles in living and passivated cells. We demonstrate that the antisymmetric part of the mean back relaxation can detect the mentioned broken time reversal symmetry.

To allow for time reversal symmetry breaking, we extend the previously introduced random horse and cart model to a non-Gaussian version, thereby introducing a length and a time scale of driving. 

Evaluating time reversal symmetry breaking via MBR for this model reveals that by changing the length and time parameters of MBR, the mentioned length and time scales of model-activity can indeed be observed. Applying the same method to cell data shows very similar features, and we determine the length scale of pronounced breaking of time reversal symmetry to be roughly $10$ to $\SI{20}{\nano \meter}$. As for time scales, there are various signatures, most prominently at around $ \SI{500}{\milli \second}$, $ \SI{100}{\milli \second}$ and $10$ to $\SI{100}{\milli \second}$. These observations represent, to our knowledge, the first demonstration of time reversal symmetry breaking with colloidal particles \textit{in vivo}. Previous studies have observed similar effects in \textit{in vitro} systems, such as those utilizing nanotubes \cite{bacanu_inferring_2023}, or for larger systems like hair bundle cells \cite{roldan_quantifying_2021}.

Treating cells by drugs reveals  a connection between the observed time reversal breaking and activity of the cytoskeleton, particularly involving microtubules and the motor protein dynein. Other studies have similarly found that anomalous diffusion is greatly reduced if microtubules are depolymerized, but not if actin is destroyed \cite{sabri_elucidating_2020}. 

To quantify breaking of time reversal symmetry, we employ a bound expression for entropy production rate, for both the model as well as for the cell data. For the model, this bound is, as expected, much smaller than the work performed by activity. We re-emphasize a natural difficulty in extracting entropy production from a reduced set of observables. This is especially so for a one dimensional trajectory, as it requires non-Gaussianity for a finite time reversal symmetry breaking. Despite these challenges, we convincingly extract entropy production from cell data, which is in astonishing qualitative agreement to previously determined effective energies.

We note another challenge in extracting entropy production from noisy data exemplified for the passive aragose: While the antisymmetric part of MBR in \cref{fig:MBR_cells} b)  fluctuates around zero for aragose, the entropy bound expression \cref{eq:s-bound} is inherently non-negative, and systematic instrument noise, like drifts will induce a finite entropy production bound, as seen in \cref{fig:sigma_bound}, where the value of aragose is finite within error. This hints at a general difficulty when employing entropy bounds with experimental (noisy) data.

There are various possibilities for future work: It will be important to   identify the effects of specific motor proteins and observing their influence on the time and length scales appearing in time reversal symmetry breaking. It may also be interesting to investigate further the observed relation between entropy production and violation of fluctuation dissipation theorem. Such relationship may indeed be motivated by the Harada-Sasa relation, which relates, for a simple Langevin equation, the integral over FDT violation to entropy production \cite{harada_equality_2005}. Finally, one may incorporate explicit modeling of the underlying Markov chains governing the active process \cite{maes_markov_2003}.

\bibliographystyle{ieeetr}
\bibliography{references}

\appendix
\section{Additional graphs from cell data}\label{app:plots}
\begin{figure*}
    \includegraphics[width=\linewidth]{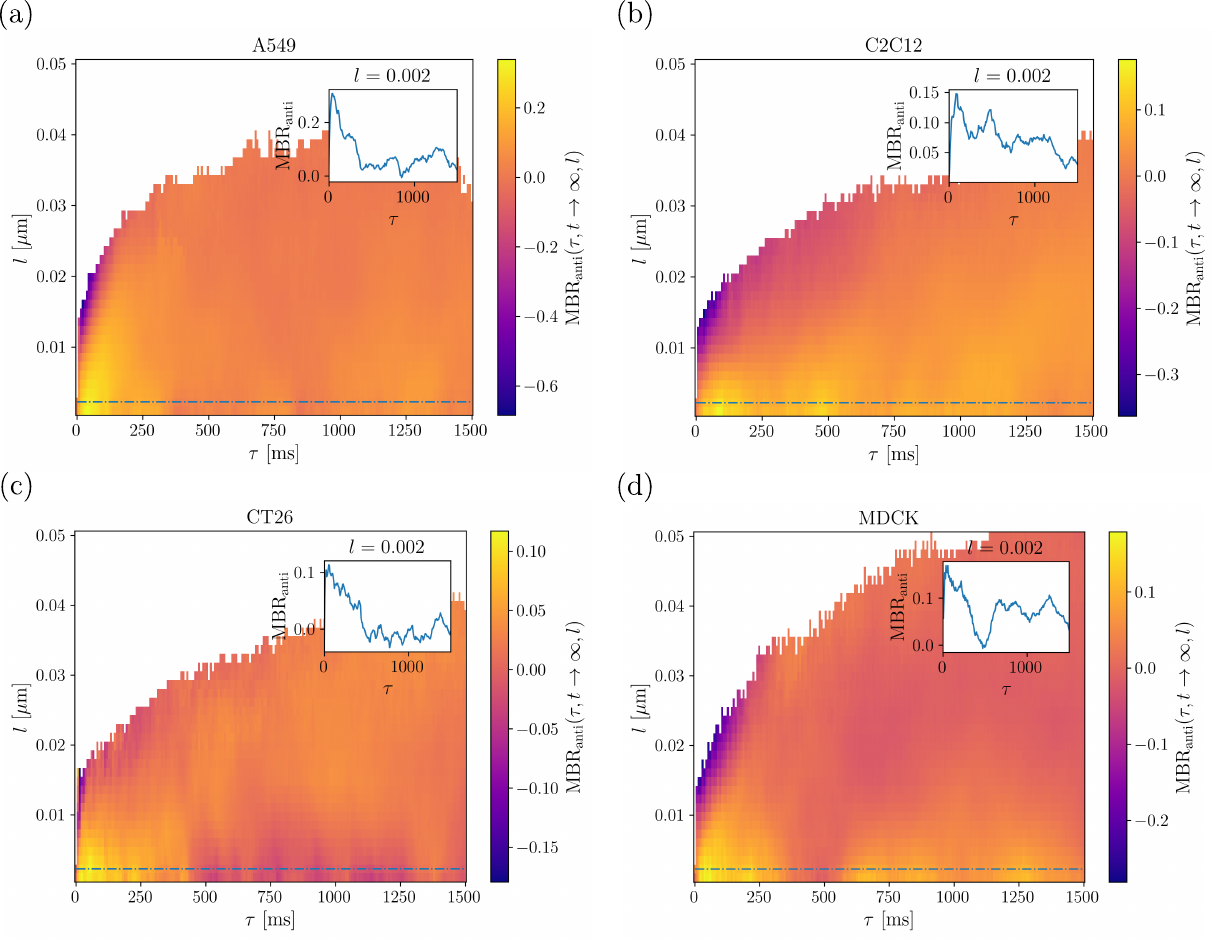}
    \caption{$\MBR_\text{anti}$ at $t=\SI{1}{\second}$ as functions of $\tau$ and $l$ as in \cref{fig:Hox-mbr-LT} for (a) A549, (b) C2C12, (c) CT26 and (d) MDCK cells. As in \cref{fig:Hox-mbr-LT} a washed out periodic structure is notable.}
    \label{fig:cells_LT}
\end{figure*}

The experimental details can be found in Ref.~\onlinecite{muenker_accessing_2024}. Multiple trajectories for different cells and media have been recorded A549 ($N=179$), C2C12 ($N=177$), CT26 ($N=173$), Drugged ($N=75$), HeLa ($N=231$), HoxB8 ($N=170$), MDCK ($N=183$), Agarose ($N=27$), HeLa CytoB ($N=60$) and HeLa Nocoda ($N=63$) with a sampling time of $1.57 \times 10^{-5}\SI{}{\second}$. For A549, the trajectory lengths lie between \SI{3}{\second} - \SI{10}{\second} with an average of \SI{8.7}{\second}. For HoxB8, the trajectory lengths lie between \SI{5}{\second} - \SI{10}{\second} with an average of \SI{5.6}{\second}. For all other cells and media, the trajectory lengths are exactly \SI{10}{\second}. The anti-symmetric MBR value at $t=\SI{1}{\second}$ for A549, C2C12, CT26  and MDCK cells can be found in \cref{fig:cells_LT}.

\end{document}